\documentclass[aps,10pt,prd,preprintnumbers,amsmath,amssymb,nofootinbib,superscriptaddress,a4paper,twocolumn,showpacs]{revtex4-1} 
\usepackage{amsmath,amsthm,amssymb}
\usepackage{bm,bbm}
\usepackage{graphicx}
\usepackage{color}
\usepackage[usenames,dvipsnames]{xcolor}
\usepackage[breaklinks]{hyperref}
\usepackage[titletoc]{appendix}

\newcommand{\be}{\begin{equation}}
\newcommand{\ee}{\end{equation}}
\newcommand{\ba}{\begin{eqnarray}}
\newcommand{\ea}{\end{eqnarray}}
\def\bs{\begin{subequations}}
\def\es{\end{subequations}}
\def\a{\alpha}
\def\b{\beta}
\def\de{\delta}
\def\g{\gamma}
\def\la{\lambda}

\def\e{\epsilon}

\def\Om{\Omega}
\def\om{\omega}

\def\s{\sigma}
\def\vr{\varrho}
\def\vp{\varphi}
\def\N{\nabla}
\def\cA{\mathcal{A}}
\def\cB{\mathcal{B}}

\def\cD{\mathcal{D}}
\def\cE{\mathcal{E}}
\def\cF{\mathcal{F}}

\def\cJ{\mathcal{J}}
\def\cK{\mathcal{K}}
\def\cL{\mathcal{L}}

\def\cQ{\mathcal{Q}}

\def\cS{\mathcal{S}}

\def\cV{\mathcal{V}}

\def\ds{d_{\rm S}}
\def\dh{d_{\rm H}}

\def\p{\partial}

\def\B{\Box}
\newcommand{\Eq}[1]{(\ref{#1})}
\def\com{\color{magenta}}
\def\cob{\color{blue}}

\newcommand{\book}[5]{\emph{#1} (#2, #3, #4, #5)}
\newcommand{\books}[4]{\emph{#1} (#2, #3, #4)}
\newcommand{\oarX}[1]{\href{http://arxiv.org/abs/#1}{{\ttfamily\com #1}}}
\newcommand{\arX}[1]{\href{http://arxiv.org/abs/#1}{{\ttfamily\com arXiv:#1}}}
\newcommand{\doin}[6]{\href{http://dx.doi.org/#1}{{\cob #2 #3 {\bf #4}, #5 (#6)}}}
\newcommand{\doinn}[5]{\href{http://dx.doi.org/#1}{{\cob #2 {\bf #3}, #4 (#5)}}}
\newcommand{\doij}[5]{\href{http://dx.doi.org/#1}{{\cob #2 #3 (#5) #4}}}

\newcommand{\ndoinn}[5]{\href{#1}{{\cob #2 {\bf #3}, #4 (#5)}}}
\newcommand{\tia}[1]{}
\newcommand{\boxd}[1]{\boxed{\phantom{\Biggl(}#1\phantom{\Biggl)}}}
\def\lp{\ell_{\rm Pl}}
\def\rme{\text{e}}
\def\rmd{\text{d}}
\def\rmi{\text{i}}

\newcommand{\fl}{}

\begin{document}

\title{Varying electric charge in multiscale spacetimes}

\author{Gianluca Calcagni}
\email{calcagni@iem.cfmac.csic.es}
\affiliation{Instituto de Estructura de la Materia, CSIC, Serrano 121, 28006 Madrid, Spain}
\author{Jo\~ao Magueijo}
\email{magueijo@ic.ac.uk}
\affiliation{Theoretical Physics, Blackett Laboratory, Imperial College, London, SW7 2BZ, United Kingdom}
\affiliation{Dipartimento di Fisica, Universit\`a La Sapienza and Sez.\ Roma1 INFN, Piazzale A.\ Moro 2, 00185 Roma, Italia}
\author{David Rodr\'iguez Fern\'andez}
\email{davidrodriguezfernandez@estumail.ucm.es}
\affiliation{Departamento de F\'isica Te\'orica II, Universidad Complutense de Madrid, 28040 Madrid, Spain}

\date{May 14th, 2013}

\begin{abstract}
We derive the covariant equations of motion for Maxwell field theory and electrodynamics in multiscale spacetimes with weighted Laplacian. An effective spacetime-dependent electric charge of geometric origin naturally emerges from the theory, thus giving rise to a varying fine-structure constant. The theory is compared with other varying-coupling models, such as those with a varying electric charge or varying speed of light. The theory is also confronted with cosmological observations, which can place constraints on the characteristic scales in the multifractional measure. We note that the model considered here is fundamentally different from those previously proposed in the literature, either of the varying-$e$ or varying-$c$ persuasion.
\end{abstract}



\preprint{\doin{10.1103/PhysRevD.89.024021}{Phys.\ Rev.}{D}{89}{024021}{2014} \hspace{10.5cm} \arX{1305.3497}}

\maketitle


\section{Introduction}

In the last few years, interest has been raised on spacetimes which, due to quantum-gravity effects, show anomalous behaviour of their geometry. Independent approaches to quantum gravity, ranging from asymptotic safety, noncommutative geometry, causal dynamical triangulations, and spin foams to fractal field theory, Ho\v{r}ava--Lifshitz and super-renormalizable gravity, display a feature known as {dimensional flow} or dynamical dimensional reduction, namely, the change of spacetime dimensionality with the scale \cite{tHo93,Car09,fra1}. The quest for a quantum theory of gravity presently aims, among other and perhaps more urgent objectives, to understand why effective quantum geometry exhibits, at short distances, the typical properties of fractals. The presence of this almost universal behaviour encouraged to seek common explanations and relations among the theories, as well as the relation between their desired-for ultraviolet (UV) finiteness and dimensional flow.

A framework where these questions may be posed with clarity is field theory in multiscale, in particular multifractional, spacetimes (short presentations of these models can be found in \cite{fra1,fra4,fra6,AIP}). This is a field theory in an ordinary sense, constituted by a set of hand-made tensor and spinor fields on a continuum obeying an action principle and the usual quantization rules, but such that the measure of the action represents a geometry with anomalous properties. The form of this measure is dictated by requiring that it encodes certain structures of multifractal geometry, including the presence of a hierarchy of scales. Modulo a change in the geometry (and, hence, the symmetries) of the model, one should be able to ask the usual questions one can answer in a conventional  perturbative field theory, thus opening up the possibility to study the above-mentioned relation between renormalizability and dimensional flow, even in the absence of gravity. However, the definition of the theory itself is still in progress and one has to start by defining simple classical systems later to be quantized. 

This program was started in concrete in \cite{frc6} for a real scalar field. Here we shall extend the treatment, only at the classical level, to an Abelian gauge field and construct first Maxwell's action and then electrodynamics with fermions. The procedure for deriving the equations of motions and the energy-momentum tensor is the one of \cite{frc6} adapted to vector and spinor fields. We shall work out classical Maxwell theory and its equations for the electric and magnetic fields, as well as the equations of motion of electrodynamics in the presence of fermions and a $U(1)$ gauge vector field. As expected, Maxwell's equations for the electric and magnetic fields are affected by the anomalous nature of the background geometry; in particular, we will find a modified conservation law for the charge density which suggests that the electric charge may vary in time and position.

Although the subject is of intrinsic interest, we must highlight an important spin-off resulting from the existence of an effective electron charge which depends on the spacetime measure. Models with spacetime-dependent couplings (violating the equivalence principle) have received considerable attention especially in the sector of electrodynamics. The fine-structure constant $\a=e^2/(\hbar c)$ depends on the electron charge $e$, the Planck's constant $\hbar$, and the speed of light $c$. When one or more of these constants are promoted to coordinate-dependent parameters, one effectively obtains a time-space varying $\a$.

Theories with varying couplings, either at an effective or a fundamental level, are not a novelty in theoretical physics. Aside from sheer scientific curiosity, they find some justification in the fact that observations, both terrestrial and astronomical \cite{ShB2,BaS1}, do not exclude that the constants of Nature are, eventually, not really constant \cite{Uza10}. For instance, there exist constraints on the variation of the fine-structure constant, which depend on the time and spatial scale of the experiment \cite{Uza10,Bar10,Chi11,Orz12}. A comparison of rates between different atomic and molecular clocks in laboratory give $|\Delta\a/\a|< 10^{-14}\div 10^{-17}$ (see \cite{Uza10,Orz12} for a detailed compilation of results), where $\Delta\a=\a(t)-\a(t_0)$ is the change in $\a$ from some time $t$ in the past with respect to today's value $\a(t_0)$. Model-dependent data analyses of the Oklo nuclear-fission event of about 1.8 billion years ago obtain roughly $|\Delta\a/\a|< 10^{-8}$ \cite{Uza10}. Further back in time, low-redshift quasars at $z=0.25$ and $z=0.68$ (corresponding to a look-back time of about $3.0$ and $6.4$ Gyr, respectively, using the best-fit parameter estimates from the Planck+WP+highL dataset \cite{Pl13a}) indicate that $|\Delta\a/\a|<10^{-5}$ \cite{DWBF,Car00,Mu01a}. Recent observations (at Keck observatory) of quasars at larger redshift yield an almost $5\s$ evidence that the fine-structure constant was smaller at early epochs \cite{MWF,Mur03,KMWM},
\be\label{deaa}
\frac{\Delta\a}{\a} = (-0.57 \pm 0.11)\times 10^{-5}\,,\qquad 0.2<z<4.2\,,
\ee
which can be further split according to the redshift range: $\Delta\a/\a = (-0.54 \pm 0.12)\times 10^{-5}$ for $z<1.8$, $\Delta\a/\a = (-0.74 \pm 0.17)\times 10^{-5}$ for $z>1.8$.

However, data from the Very Large Telescope (VLT) on different samples and using different methods of analysis gave results compatible with a nonvariation of $\a$ \cite{CSPA,SCPA1}, and there is some ongoing debate \cite{MWF06,MWF07,SCPA2} (see \cite{Uza10} for a review). To make things even more confusing, further data from VLT seems again to indicate a variation of $\a$ at large redshift, but of opposite sign with respect to Eq.\ \Eq{deaa} \cite{Web10,Kin12},
\be\label{deaa2}
\frac{\Delta\a}{\a} = (+0.21 \pm 0.12)\times 10^{-5}\,,\qquad 0.2<z<3.7\,,
\ee
again split in two estimates: $\Delta\a/\a= (-0.06 \pm 0.16)\times 10^{-5}$ for $z<1.8$ and  $\Delta\a/\a = (+0.61 \pm 0.20)\times 10^{-5}$ for $z>1.8$. This can be reconciled with Eq.\
 \Eq{deaa} via a dipole model \cite{Web10,Kin12} if $\a$ admitted spatial variations (Keck and VLT are located in different hemispheres and probe different directions in the sky).
Other quasar observations do not show a temporal change in $\alpha$ ($\Delta\a/\a \lesssim 10^{-6}$ at $z\sim 1.7$) but they are not incompatible with the dipole model either \cite{Mol13}. Still, the detection of a varying $\a$ and a dipole effect is controversial and unconfirmed (or contradicted) by yet other quasar data ($\Delta\a/\a < 10^{-6}$ at $z\sim 1.3$) \cite{Rah12}, lensed galaxy spectra ($\Delta\a/\a < 10^{-5}$ at $z\sim 5.2$) \cite{Lev12} and data reanalyses \cite{CaP}. Spectra observations of nearby ($\sim 45$ pc) white dwarfs may provide independent constraints on the variation of $\alpha$ in the near future \cite{Ber13}.

Any varying-$\a$ field theory should be able to comment on these findings, either by explaining the dipole effect or, through the above experiments, placing constraints on the free parameters (if any) governing the spacetime dependence of the fine-structure parameter $\a$. The latter is one of the goals of this paper.

The plan is as follows. In Sec.\ \ref{revmf}, we review the theory of multiscale spacetimes. Then, in Sec.\ \ref{em} we set up a gauge-invariant electromagnetic theory living on such a space. In Sec.\ \ref{comp}, we compare our framework with varying-$e$ and varying-$c$ models proposed in modern and contemporary literature, commenting on how the phenomenology of multifractional spacetimes fares with respect to these approaches and the above physical constraints. Finally, in a concluding section we summarize our results and outline future work. 


\section{Brief review of multiscale spacetimes with weighted Laplacian}\label{revmf}

Field theories in multiscale spacetimes are defined by an action
\be\label{acti}
S=\int_{-\infty}^{+\infty} \rmd\vr(x)\mathcal{L}\,,
\ee
where $\vr(x)$ is a generic Lebesgue--Stieltjes measure with anomalous scaling.  We assume that the measure can be written as $\rmd\vr(x)=\rmd^Dx\,v(x)$, i.e., is the usual $D$-dimensional volume element multiplied by a distribution law $v(x)$, where $D$ is the number of topological dimensions. To make the problem tractable, this measure weight should have the following properties:
\begin{enumerate}
\item Be factorizable in the coordinates,
\be
v(x)=\prod_{\mu=0}^{D-1}v_\mu(x^\mu)\,,
\ee
where the $D$ functions $v_\mu$ may be all different. In the ``isotropic'' case, they are all equal.
\item Be positive semidefinite, $v_\mu =v_\mu (x)\geq 0$.
\end{enumerate}
Violation of either of these conditions would hinder the definition of an invertible momentum transform \cite{frc3} and the construction of Noether currents \cite{frc6}, and would give rise to other problems at the level of quantum mechanics \cite{frc5}.

The prototypical measure obeying a neat anomalous scaling law is fractional, i.e., of the form
\be\label{frame}
v(x)=v_\a(x)=\prod_{\mu}v_\a(x^\mu):=\prod_{\mu=0}^{D-1}\frac{|x^\mu|^{\a_\mu -1}}{\Gamma(\alpha_\mu)}\,,
\ee
where $0< \alpha_\mu \leq 1$ are $D$ parameters (all equal to their average $\a:=(\sum_\mu \a_\mu)/D$ in the isotropic case) and the factor $\Gamma(\alpha_\mu)$ is inherited from the definition of fractional integral associated with this kind of measures \cite{frc1}. It is easy to check that the Hausdorff dimension of spacetimes endowed with this class of measures is given by $\vr(\la x)=\la^{\dh}\vr(x)$, where $\dh=D\a\leq D$, as can be found also by looking at the way balls scale with the radius \cite{frc6,frc1}. Fractional measures of the form \Eq{frame} have been shown (in a one-dimensional embedding) to represent random fractals as well as an approximation of deterministic fractals \cite{RYS,YRZ,RYZLM,Yu99,QL,RQLW,RLWQ,NLM} (see \cite{frc1} for a discussion on this approximation). In this precise sense, the choice of \Eq{frame} is in direct contact with fractal geometry and singled out among all possible arbitrary functional profiles $v(x)$.

To get a geometry where the spacetime dimension varies with the probed scale, it is sufficient to sum over a minimum of two values of $\a$ \cite{frc6,frc2,frc4}:
\ba
v(x) &=&v_*(x)=\prod_\mu v_*(x^\mu)\nonumber\\
&:=&\prod_\mu\left[\sum_{n=1}^N g_{\mu,n}(\{\ell_n^\mu\}_n)\, v_{\a_n}(x^\mu)\right]\!,\label{muf2}
\ea
where $N$ is integer and the dimensionful couplings $g_{\mu,n}$ depend on a hierarchy of length scales $\ell_n^\mu$. When $N=2$, the measure is called binomial and it realizes a monotonic dimensional flow between two asymptotic regimes. In particular, to get $\dh=D$ in the infrared, one can take the spatially isotropic choice 
$\a_{0,1}=\a_0$, $\a_{i,1}=\a_*<1$, $\a_{\mu,2}=1$, $g_{0,1}=\Gamma(\a_0) |t_*|^{1-\a_0}$, $g_{i,1}=\Gamma(\a_*)\ell_*^{1-\a_*}$, and $g_{\mu,2}=1$, where $\ell_*$ is a fundamental spatial length and $t_*$ a characteristic time. Then, in the UV $\dh=\a_0+(D-1)\a_*$ and
\be\label{binom}
v_*({\bf x})=\prod_{i=1}^{D-1}\left[1+\left(\frac{|x^i|}{\ell_*}\right)^{\a_*-1}\right]\!,\quad v_*(t)=1+\left|\frac{t}{t_*}\right|^{\a_0-1}.
\ee
Further generalizations, as to log-oscillating measures, are possible \cite{fra4,frc2,ACOS} and even in closer contact with fractal geometry \cite{NLM}.

The Lagrangian density $\cL$ in the action \Eq{acti} is made up of tensor fields as in theories in ordinary spacetime, except that the differential structure of the geometry they live on is modified by the nontrivial measure. There exist various models of multifractional geometry depending on the symmetries of the Lagrangian (see \cite{frc7} for a detailed discussion), but here we choose to consider the one where the natural derivative is a self-adjoint operator with respect to the scalar product with measure $\vr$, so that the Laplace--Beltrami operator $\B=\p_\mu\p^\mu$ of Minkowski spacetime $M^D$ is replaced by
\be
\cK_v:=\eta^{\mu\nu}\cD_\mu \cD_\nu\,,\qquad \cD_\mu:=\frac{1}{\sqrt{v(x)}}\p_\mu \left[\sqrt{v(x)}\,\cdot\,\right],
\ee
where $\eta={\rm diag}(-,+,\cdots,+)$ is the Minkowski metric. Thus, when constructing a field theory on these spaces, it is natural (but not sufficient, as we will see below) to take the standard action for the fields of interest (scalars, vectors, fermions, and so on) and make the replacements $\rmd^Dx\to\rmd\vr(x)$ and $\p_\mu\to\cD_\mu$.

With this Laplacian, one can then derive the diffusion equation through a Langevin-equation approach and, from that, the spectral dimension $\ds$ of spacetime, which is anomalous in general (that is, $\ds\neq\dh\neq D$) \cite{frc7}.


\section{Electromagnetism and fermions in multifractional field theory}\label{em}

In this section, we set up electromagnetism in the spacetime reviewed just above. We note that the topic of electrodynamics in multifractional spacetimes is practically virgin. To the best of our knowledge, a Maxwell field theory has been briefly considered only in \cite{MSBR}, for a fractional measure and an ordinary Lagrangian. There are some results in other, quite different models of anomalous spacetimes, such as the one due to Stillinger \cite{Sti77,PSt} (where the electromagnetic wave equation has been studied \cite{ZMN1,ZMNR,ZMN2,ZMN3,ZMN4}) and Nottale's scale relativity \cite{Not93,Not96,Not08} (Maxwell action and electrodynamics \cite{Not94,Not03,NCL}). Both approaches have been compared with the multifractional framework in \cite{frc2,fra7}. Fractional calculus has been more extensively employed to describe Maxwell electrodynamics with fractal charge distributions \cite{Ta05a,Ta05b} or in fractal turbulent media \cite{Tar06,Tar8} (where a fractional integration measure makes its appearance), or in dielectric media and various problems \cite{Eng95,Eng96,Eng97,Eng98,Eng99,Lak01,VeE,IV,NaA,NNH,HN,HIN,MuB,Tar13,Tar12,BGG,HNIN} (where effective electromagnetic equations sport fractional derivatives). These scenarios are neither associated with intrinsically anomalous spacetimes (except as a heuristic motivation for \cite{MuB,BGG}) nor formulated in a field-theory context.


\subsection{Maxwell action and equations}

In this subsection only, we will fix the units so that $c=1=\hbar$. The equations of motion are derived from a variational principle $\de S=0$,\footnote{A fractional version $\de_v$ of functional variations exists \cite{frc6}, but in the present setting it gives the same equations of motion.} where $\de$ represents the field variation at a given point, $\de f:=f'(x)-f(x)$ for any field $f$ (hence, $v$ being a fixed coordinate profile, $\de(vf)=v\de f$). Assuming that the Lagrangian density only depends on an Abelian gauge field $A_\mu$ and its weighted derivatives $\cD_\mu A_\nu$, one has (integration is over the whole embedding $M^D$)
\ba
\de S &=&\int \rmd^Dx\,v(x)\de\mathcal{\cL}(A_\nu, \cD_{\mu}A_\nu)\nonumber\\
      &=&\int\rmd^Dx\,v(x)\left[\frac{\p\cL}{\p A_\nu}\de A_\nu+\frac{\p\cL}{\p(\cD_\mu A_\nu)}\de(\cD_\mu A_\nu)\right]\nonumber\\
      &=&\int\rmd^Dx\,v(x)\left[\frac{\p\cL}{\p A_\nu}\de A_\nu+\frac{\p\cL}{\p(\cD_\mu A_\nu)}\cD_\mu\de A_\nu\right],
\ea
where we used the field-variation property $[\cD_\mu,\de]A_\nu=0$. Using now the Leibniz rule
\be
\cD_\mu (ab)=(\p_\mu a)b+a\cD_\mu b\,,
\ee
it is not difficult to show that
\ba
\hspace{-1cm}\de S &=& \int\rmd^Dx\,v(x)\left\{\frac{\p\cL}{\p A_\nu}\de A_\nu -\frac{1}{2}\frac{\p_\mu v}{v} \frac{\p \cL}{\p (\cD_\mu A_\nu)}\de A_\nu\right.\nonumber\\
&& \left. -\p_\mu\left[\frac{\p\cL}{\p(\cD_\mu A_\nu)}\right]\de A_\nu+\p_\mu \left[\frac{v\p \cL}{\p (\cD_\mu A_\nu)}\de A_\nu\right]\right\}\!.\label{tode}
\ea
The last term is a total derivative and it can be thrown away by virtue of the condition $\de A_\mu\to 0$ on the field variation, when $x^\mu\to\pm\infty$ (later on, we will keep it when computing the variation under a coordinate transformation). Also, a continuity condition \cite{frc6} guarantees that the boundary term is zero at $x^\mu=0^\pm$. Thus, the equations of motion read
\be\label{eom}
\frac{\p\cL}{\p A_\nu}-\cD_\mu \left[\frac{\p\cL}{\p(\cD_\mu A_\nu)}\right]=0\,.
\ee
Notice that the same form of Euler--Lagrange equations (as well as the ensuing conservation laws below) would hold for a derivative with arbitrary weight,
\be\label{bcD}
\cD\to {}_\b\cD:=\frac{1}{v^\b}\p[v^\b\,\cdot\,]\,.
\ee
However, only for the self-adjoint case $\b=1/2$ is the theory invariant under a reordering of the operators in the action.

\subsubsection{Maxwell equations}

Let us now choose $\cL$ to be the Maxwell Lagrangian with source $J$,
\ba
\cL &=& \cL_F+ J_\mu A^\mu\,,\label{maxL}\\
\cL_F &=& -\frac1{4} F_{\mu\nu} F^{\mu\nu}\,,\label{lf}
\ea
where
\be\label{maxf}
F_{\mu\nu}:=\cD_\mu A_\nu-\cD_\nu A_\mu
\ee
is the (antisymmetric) field strength of the gauge vector. Then, one obtains the Maxwell equations with source,
\be\label{max1}
\cD_\nu F^{\mu\nu}=J^\mu\,.
\ee
Applying the operator $\cD_\mu$ to this equation, and taking into account that $[\cD_\mu,\cD_\nu]=0$ for factorizable measures, we find that the current $J$ obeys the (non-)conservation law
\be\label{dj}
\cD_\mu J^\mu=0\,.
\ee
Thus, the action \Eq{acti} with Lagrangian density \Eq{maxL} is invariant under the gauge transformation
\be\label{gauge}
A_\mu\to A_\mu+\cD_\mu\phi\,,
\ee
where $\phi$ is a scalar field \emph{density}.

In $D=4$ dimensions, writing down the field strength in terms of its components,
\be
F^{\mu\nu}=\begin{pmatrix}
						 0   & E_1  & E_2  & E_3\cr
						-E_1 & 0	  & B_3  & -B_2\cr
						-E_2 & -B_3	& 0    & B_1\cr
						-E_3 & B_2  &	-B_1 & 0\cr
						\end{pmatrix}\,,
\ee
we get the first pair of Maxwell equations from \Eq{max1}, telling how the divergence of the electric field and the curl of the magnetic field depend on the source:
\be\label{max1b}
\cD\cdot {\bf E}=\cD_i E^i=J^0\,,\qquad \cD\times {\bf B}-\cD_t {\bf E} = {\bf j}\,,
\ee
where the index $i$ runs on spatial directions and $j^i=J^i$. In ordinary spacetime, the component $J^0=\rho$ is the charge density and, integrating the first equation over the whole volume, one gets the total charge, proportional to the electron charge $e$, which is a constant. To see what we should expect instead in a multifractional ambient space, we notice that Eq.\ \Eq{dj} leads to 
\be\label{rhoj}
\cD_t\rho+\cD\cdot {\bf j}=0\,.
\ee
Thus, even in the absence of spatial current density, the charge density $\rho$ is \emph{not} conserved in time, $\dot\rho =-\rho\dot v /(2v)$. In general, from the time component of the current density $J^\mu$ one defines the electric charge
\be\label{qch}
Q:=\int\rmd\vr({\bf x})\,J^0
\ee
as an integral over the multifractional spatial volume of the charge density. Applying the operator $\cD_t$ or $\p_t$ to the $\mu=0$ component, one finds $\cD_t Q\neq0\neq \dot Q$, due to the fact that the left weight $v^{-1/2}$ in $\cD_i$ in \Eq{rhoj} does not cancel the one in the spatial measure, and one does not obtain a total divergence. 

The source of this novelty lies in the fact that $J$ is a vector density with weight $-1/2$ with respect to $v$.
Due to the nontrivial weight $v$ in the action \Eq{acti}, $\cL$ is a density, and so are $A_\mu$, $J^\mu$ (vector densities), and $F_{\mu\nu}$ (rank-2 tensor density). This immediately defines actual tensorial quantities $\cA_\mu$, $\cJ^\mu$, and $\cF_{\mu\nu}$ under the field redefinition
\be
\cA_\mu :=\sqrt{v}\,A_\mu,\quad \cF_{\mu\nu}[\cA]=\p_\mu \cA_\nu-\p_\nu \cA_\mu=\sqrt{v} F_{\mu\nu}[A],\label{FcF}
\ee
\be
\cJ^\mu:=\sqrt{v}\,J^\mu\,,\label{FcF2}
\ee
so that, in particular, the Maxwell action can be reduced to the ordinary one:
\be
S[A,J]=\cS[\cA,\cJ]=\int\rmd^Dx\, \left(-\frac14\cF_{\mu\nu} \cF^{\mu\nu}+\cJ_\mu \cA^\mu\right).
\ee
Notice the complete disappearance of the measure weight into the new fields. This is nothing but the ``integer picture'' of \cite{frc6}, such that free fractional systems can be formally (the geometry is still anomalous) mapped onto ordinary systems via a field redefinition. The deduction of the equations of motion \Eq{max1} could have proceeded as in the standard case by working with $\cA$ and $\cJ$.

There are three remarks one can make here. First, whilst under an ordinary Lorentz transformation ${\cA'}^\mu(x')=\Lambda^\mu_{\ \nu}\cA^\nu(x)$ transforms as a vector, the density $A$ acquires an extra weight factor:
\be\label{La}
{A'}^\mu(x')=\sqrt{\frac{v(x)}{v(x')}}\Lambda^\mu_{\ \nu} A^\nu(x)\,,
\ee
where 
\be\label{lore}
{x'}^\mu=\Lambda^\mu_{\  \nu} x^\nu
\ee
and $\Lambda^\mu_{\ \nu}$ is the usual Lorentz transformation matrix. The vector density ${A'}^\mu=\Lambda_\om A^\mu$ is defined via the operator
\be\label{Uomega}
\Lambda_\omega := \rme^{-\frac{\rmi}2 \omega^{\mu \nu} \Om_{\mu \nu}}=\frac{1}{\sqrt{v}}\,\bar \Lambda_\omega\,\sqrt{v}\,,\qquad \bar \Lambda_\omega := \rme^{-\frac{\rmi}2 \omega_{\mu \nu} \bar\Om^{\mu \nu}},
\ee
where $\omega$ is an antisymmetric matrix of parameters and $\Om$ are the fractional Lorentz transformations in $D$-vector representation:
\be 
\Om^{\nu\rho}:= \frac{1}{\sqrt{v}}\,\bar\Om^{\nu\rho}\,\sqrt{v}\,,\qquad \left(\bar\Om^{\nu\rho}\right)^\mu_\s=\rmi(\eta^{\mu\rho}\de^\nu_\s -\eta^{\nu\rho}\de^\mu_\s)\,.
\ee
The $\bar\Om^{\nu\rho}$ generate ordinary rotations and boosts. Expanding as $\Lambda_\om \approx \mathbbm{1}-(\rmi/2)\omega_{\mu \nu} \Om^{\mu \nu}$, $\sqrt{v(x)/v(x')}\approx 1-\p_\mu v \de x^\mu/(2v)$, and noting that $\de x^\mu=-(\rmi/2)\om_{\rho\s}(\bar\Om^{\rho\s})^\mu_{\ \nu} x^\nu$, it is easy to find the infinitesimal version of Eq.\ \Eq{La}. The infinite-dimensional field representations of the fractional Lorentz generators can be derived along similar lines \cite{frc6}.

Second, assuming $D=4$, $\cF^{0i}=\cE^i$ and $\cF^{ij}=\e^{ijk}\cB_k$ obey the ordinary Maxwell equations $\N\cdot {\bf \cE}=\bar \rho$ and $\N\times {\bf \cB}-\dot {\bf \cE}=\bar {\bf \jmath}$, with
\be\label{baro}
\bar\rho(x):=\cJ^0=\sqrt{v(x)}\rho(x)
\ee
and $\bar \jmath^i=\cJ^i$. From the divergence of the current \Eq{dj}, $\p_\mu \cJ^\mu=0$, one has charge conservation, $\dot\cQ=0$, where
\be\label{qch2}
\cQ:=\int\rmd{\bf x}\,\cJ^0\,.
\ee
One can directly check conservation of $\bar\rho$ by integrating Eq.\ \Eq{rhoj} in $\int\rmd{\bf x}\sqrt{v({\bf x})}$.
The sourceless Maxwell equations are immediately given by the Bianchi identity for the antisymmetric 2-tensor $\cF$: $\p_{\mu_1}\tilde\cF^{\mu_1\cdots\mu_{D-2}}=0$, where $\tilde\cF^{\mu_1\cdots\mu_{D-2}}= \e^{\mu_1\cdots\mu_D}\cF_{\mu_{D-1}\mu_D}/2$ is the dual field strength and $\e^{\mu_1\cdots\mu_D}$ is the Levi-Civita symbol in $D$ dimensions. In $D=2$, $\cF$ and its dual are proportional to each other (there is only one nonvanishing component) but the theory is trivial since there are no propagating degrees of freedom. In $D=3$ there is only one sourceless equation, and the missing one ($\N\cdot \cB=0$) signals the possibility of magnetic vortices. Only in $D=4$ do the number of equations with and without source coincide. In general, the Bianchi identities yield $D(D-1)/2$ independent equations. All this applies also to the multifractional model, where the Bianchi identity and Eq.\
 \Eq{FcF} give
\be\label{max2}
\cD_{\mu_1} \tilde F^{\mu_1\cdots\mu_{D-2}}=0\,,\quad \tilde F^{\mu_1\cdots\mu_{D-2}}= \frac12\e^{\mu_1\cdots\mu_D}F_{\mu_{D-1}\mu_D}\,.
\ee
In particular, in four dimensions it corresponds to
\be\label{max2b}
\cD\cdot {\bf B}=\cD_i B^i=0\,,\qquad \cD\times {\bf E}+\cD_t{\bf B}=0\,.
\ee

The third consequence of having an anomalous geometry is that, looking at Eq.\ \Eq{baro}, one is led to define an effective spacetime-dependent electron charge,
\be\label{ev}
e_v(x):=\frac{e_0}{\sqrt{v(x)}}\,,
\ee
assuming a given volume. In fact, for a uniform charge distribution $\bar\rho= n e_0/\cV_1$, while $\rho=n e_v/\cV_v$, where $n$ is the number of charges in the volume. If the two volumes were equated numerically, one would obtain the relation \Eq{ev}. However, in fact they carry different measure weights, since $\cV_v/\cV_1\sim v$, and it would be more natural to define another fractional charge $\tilde e\sim\sqrt{v} e_0$. This charge will indeed appear in Sec.\ \ref{fermions}. For the time being, we discuss Eq.\ \Eq{ev} and its relation with the observed electric charge.

Notice that the Maxwell field strength \Eq{maxf} can then be written as
\be\label{myF}
F_{\mu\nu}=e_v[\p_\mu(e_v^{-1}A_\nu)-\p_\nu(e_v^{-1}A_\mu)]\,,
\ee
later to be compared with Eq.\ \Eq{bekac}. Equation \Eq{ev} will emerge later on in a slightly different way, but we must already mention a caveat associated with it. The quantity $e_v$ is \emph{not} the charge sourcing the physical electric field $E$ living in multifractional spacetime, since $Q\neq \cQ/\sqrt{v}$, and $e_v$ does not represent the fractional electric charge $Q$ measured in a given Hausdorff volume. So which quantity can one measure in an experiment? To answer this question, we rewrite $Q$ via \Eq{FcF2} as
\ba
Q&=&\int\rmd{\bf x}\,v({\bf x})\,J^0=\int\rmd{\bf x}\,v({\bf x})\,\frac{\cJ^0}{\sqrt{v({\bf x}) v_0(t)}}\nonumber\\
&=&\frac{1}{\sqrt{v_0(t)}}\int\rmd{\bf x}\,\sqrt{v({\bf x})}\,\cJ^0\,.\label{qch3}
\ea

At this point, we recall that in multiscale spacetimes dimensional flow is possible because a tower of scales is established, where the top corresponds to the scale measured by a macroscopic observer. This happens not only in multifractional spacetimes but also in other field-theory approaches to quantum gravity, in particular asymptotic safety \cite{fra7}. Therefore, it would be desirable to express Eq.\ \Eq{qch3} in terms of the electric charge $\cQ$ one would measure in a world with integer, ordinary geometry. To this purpose, we plug the binomial measure \Eq{binom} in Eq.\ \Eq{qch3} and expand the factor $\sqrt{v_*({\bf x})/v_*(t)}$ in the limit of small $\ell_*$:
\be\label{qch4}
Q\approx \frac{1}{\sqrt{v_*(t)}}\int\left[\prod_{i=1}^{D-1}\rmd x^i\,\left(1+\frac12 \left|\frac{x^i}{\ell_*}\right|^{\a-1}\right)\right]\cJ^0\,.
\ee

To leading order, one ends up with
\be\label{qch5}
Q\approx \frac{\cQ}{\sqrt{v_*(t)}}\,,
\ee
which identifies an effective time-dependent observed charge
\be\label{e*}
\boxd{e_*(t):=\frac{e_0}{\sqrt{v_*(t)}}\,.}
\ee
In particular, at times $t\gg t_*$,
\be\label{e*t}
e_*(t)\sim \left(1-\frac12\left|\frac{t_*}{t}\right|^{1-\a_0}\right)e_0\,,
\ee
which is positive definite consistently with the $t_*/t$ expansion. The result \Eq{e*} is actually exact for a uniform charge distribution, because the constant term of the spatial measure weight always dominates in the numerator of Eq.\ \Eq{qch3} at large $x^i$ (since $0<\a<1$), and it cancels the denominator in $\cJ^0=\cQ/(\int \rmd{\bf x})$. Inclusion of spatial gradients or, more generally, an inspection of the system at small scales will modify the observed charge to a hybrid between Eq.\ \Eq{ev} and \Eq{e*}, via \Eq{qch3}.

A comment here is in order. The integer picture is only a mathematical tool to simplify the problem, and one should not
reach the conclusion that multiscale theory is physically equivalent to the usual one under the field transformation \Eq{FcF}-\Eq{FcF2}. The reason is that the most general multiscale action with interacting fields is not equivalent to the usual one, due to the presence of effectively spacetime-dependent couplings after the field redefinition \cite{frc6}. Thus, it is not always possible to reabsorb all measure factors and there typically is a nontrivial geometric effect. One could, in turn, assume that the integer picture is the physical one, with the consequence that the physically observed charge is $\cQ$, not $Q$, and electromagnetism is essentially the usual one. 

This problem of ``frame choice'' (fractional versus integer) is reminiscent of the situation one encounters in Brans--Dicke theory, where one can choose between the Einstein and the Jordan frame. In the former, Einstein's equations are valid (with a constant $G$), but matter is nonminimally (but universally) coupled to a scalar field, providing a fifth force satisfying the most basic form of the weak equivalence principle. But Jordan's frame, where matter is minimally coupled, is indeed the frame where matter follows the geodesics of a metric (a stronger version of the weak equivalence principle). In Jordan's frame a varying $G$ is more evident, but in either frame we cannot deny that we have a theory with a varying coupling constant (gravitational in this case). 

A similar phenomenon was already found and discussed in the context of varying-speed-of-light theories, as we will see later. Here something similar happens: the integer picture produces the same type of effect one gets when transforming from the Jordan to the Einstein frame. In the fractional picture, one is considering a field theory on a multiscale spacetime where geometry (volumes, and so on) is measured with respect to the integro-differential structure determined by the weight $v(x)$. There, experiments entail the field densities ${\bf E}$, ${\bf B}$, the charge $Q$, and so on. The integer picture, on the other hand, is a theory with ordinary Maxwell fields ${\bf \cE}$ and ${\bf \cB}$ and measured charge $\cQ$, where geometry is standard, Maxwell theory is standard, but a spacetime dependence arises in the couplings of other sectors of the full action.

\subsubsection{Energy-momentum tensor}

To calculate the Noether current, we take the variation $\de_0 S$ with respect to a coordinate transformation $x^\mu\to x^\mu+\de x^\mu$, and denote the coordinate variation of a field $f$ as $\de_0 f:=f'(x')-f(x)$. For infinitesimal transformations and after Taylor expanding, there follows a relation between $\de_0$ and the field variation $\de$: $\de_0 f(x)= \de f(x)+\de x^\mu\p_\mu f(x)$. In \cite{frc6}, it was shown that a fractional scalar density field $\phi$ transforms as $\de_0\phi=-[\p_\mu v/(2v)]\de x^\mu \phi$ ($\neq 0$, contrary to a proper scalar) and $\de\phi=-\de x^\mu\cD_\mu\phi$. For the vector density $A_\nu$, a similar calculation from Eq.\ \Eq{FcF} yields, under a translation $\de x^\mu=-\e^\mu={\rm const}$, $A'_\nu(x)=\cA_\nu(x+\e)/\sqrt{v(x)}=\sqrt{v(x+\e)/v(x)}\, A_\nu(x+\e)$, which in infinitesimal form reads
\be\label{dea}
\de A_\nu=\e^\mu\cD_\mu A_\nu\quad\Rightarrow\quad \de_0 A_\nu= \frac12\e^\mu\frac{\p_\mu v}{v}A_\nu\,.
\ee

Consider now the total variation of the functional action under an infinitesimal coordinate transformation, $\de_0 S=\int\rmd^Dx[\de_0(v\cL)+v\cL\de_0(\rmd^Dx)]$. Since $\de_0(\rmd^Dx)=\rmd^Dx' -\rmd^Dx=\det(\p x'^\mu/\p x^\nu)\rmd^Dx-\rmd^Dx=\det(\de^\mu_\nu+\p_\nu \de x^\mu) \rmd^Dx-\rmd^Dx$ and $\det(\mathbbm{1}+O)=1+{\rm tr}(O)$ for any operator $O$, it follows that $\de_0 (\rmd^Dx)=\p_\mu \de x^\mu \rmd^Dx$ as usual. Furthermore, $\de_0(v\cL)=\de(v\cL)+\p_\mu(v\cL)\de x^\mu$, and using the equations of motion we get, for a translation,
\ba
\de_0 S &=& \int\rmd^Dx\, \left\{v\cL\p_\mu \de x^\mu+\p_\mu(v\cL)\de x^\mu\vphantom{\frac12}\right.\nonumber\\
        && \qquad\left.+\p_\mu \left[\frac{v\p \cL}{\p (\cD_\mu A_\s)}\de A_\s\right]\right\}\nonumber\\
   			&=& \int\rmd^Dx\, \p_\mu \left[v\cL \de x^\mu+\frac{v\p \cL}{\p (\cD_\mu A_\s)}\de A_\s\right]\nonumber\\
   			&\ \stackrel{\text{\tiny \Eq{dea}}}{=}\ & \int\rmd^Dx\, \p_\mu \left[-v\cL \e^\mu+\frac{v\p \cL}{\p (\cD_\mu A_\s)}\e^\nu\cD_\nu A_\s\right]\nonumber\\
        &=& -\int\rmd^Dx\, \p_\mu \left(v T^\mu_{\ ~ \nu}\right)\e^\nu\,,\label{sti}
\ea
where the last term in the first line is the total derivative in \Eq{tode} and we defined the energy-momentum tensor density,
\be\label{emt}
T_{\mu\nu} :=\eta_{\mu\nu}\cL-\frac{\p \cL}{\p (\cD^\mu A_\s)}\cD_\nu A_\s\,.
\ee

In the last line of Eq.\ \Eq{sti} we used the constancy of $\e$, which is an arbitrary parameter. For the Maxwell Lagrangian, one has
\be\label{emtemp}
T_{\mu\nu}=-\frac{1}{4}F^{\s\tau} F_{\s\tau}\eta_{\mu\nu} +F_\mu^{~\s}\cD_\nu A_\s +J^\s A_\s\eta_{\mu\nu}\,.
\ee

This equation is not gauge invariant, since it depends explicitly on $A$ and $J$. For later use, we only rewrite the second term in the right-hand side, ignoring the source contributions. As in the usual case, we can add the divergence of a rank-3 tensor (density) antisymmetric in the first two indices, $T_{\mu\nu}\to T_{\mu\nu}+\check{\cD}^\s X_{\s\mu\nu}$, where
\be
\check{\cD}_\mu:=\frac1v \p_\mu \left[v\,\cdot\,\right]\,.
\ee

In fact, $\check{\cD}^\mu\check{\cD}^\s X_{\s\mu\nu}=0$, which does not change the (non-)conservation law we will find in a moment. Also, for $\mu=0$ one has a total spatial divergence $\check{\cD}^\s X_{\s 0\nu}=\check{\cD}^i X_{i0\nu}$ which does not affect the momentum $P^\nu:= \int \rmd\vr({\bf x})\, T^{0\nu}$. Choosing $X_{\s\mu\nu}=F_{\s\mu}A_\nu$ and using Eq.\
 \Eq{max1}, one can rewrite the energy-momentum tensor \Eq{emtemp} as
\ba
T_{\mu\nu}&=&-\frac{1}{4}F^{\s\tau} F_{\s\tau}\eta_{\mu\nu} +F_\mu^{\ \s}F_{\nu\s}-J_\mu A_\nu +J^\s A_\s\eta_{\mu\nu}\nonumber\\
&=&{}^{(F)}T_{\mu\nu}+{}^{(J)}T_{\mu\nu}\,,\label{emt2}
\ea
where we split the first and last two terms in separate contributions.

The fractional Maxwell action is not invariant under translations, due to the source term. In fact, the latter is not a field but a coordinate vector profile. In particular, $\de_0 J^\mu=J^\mu(x-\e)-J^\mu(x)=-\e^\nu\p_\nu J^\mu (x)$ and
\ba
&&\int \rmd^D x'\,v(x')\cL_J'(x')-\int \rmd^D x\,v(x)\cL_J(x) \nonumber\\
&&          \qquad= \int \rmd^D x[v(x-\e)J^\mu(x-\e) A_\mu(x-\e)\nonumber\\
&&\qquad\quad-v(x)J^\mu(x) A_\mu(x)]\nonumber\\
&&          \qquad= \int \rmd^D x\left(-J^\mu A_\mu\e^\nu\p_\nu v+v J^\mu\de_0 A_\mu+v A_\mu \de_0 J^\mu\right)\nonumber\\
&&          \qquad \stackrel{\text{\tiny \Eq{dea}}}{=} \int \rmd^D x\left(-J^\mu A_\mu\e^\nu\p_\nu v+\frac12J^\mu A_\mu\e^\nu\p_\nu v\right.\nonumber\\
&&\qquad\quad \left.\vphantom{\frac12}-v A_\mu\e^\nu \p_\nu J^\mu\right)\nonumber\\
&&          \qquad= -\int \rmd^D x\,v\left(A_\mu \cD_\nu J^\mu\right)\e^\nu\,.\nonumber
\ea

Therefore, from Eq.\ \Eq{sti} and the arbitrariness of $\e$, one gets the conservation law
\be\label{Tcon}
\check{\cD}_\mu T^\mu_{~ \nu}=A_\mu \cD_\nu J^\mu\,.
\ee

The natural derivative acting on the energy-momentum tensor has weights $v$ because $T$ is a bilinear density.

The conservation law \Eq{Tcon} can be verified directly from Eq.\ \Eq{emt2}. Noting that $\check{\cD}_\tau F_{\mu\nu}+\check{\cD}_\mu F_{\nu\tau}+\check{\cD}_\nu F_{\tau\mu}=0$ and using Maxwell's equations, we have $\check{D}_\mu\, {}^{(F)}T^\mu_{\ ~\nu}=-J^\mu F_{\nu\mu}$. On the other hand, $\check{D}_\mu\, {}^{(J)}T^\mu_{\ ~\nu}=J^\mu F_{\nu\mu}+A_\mu\cD_\nu J^\mu$. Combining the two contributions yields Eq.\ \Eq{Tcon}.

The source term in Eq.\ \Eq{Tcon} arises because the action does not include the matter contribution. It is well known in standard electromagnetism that, when a charged relativistic particle is added, the total energy-momentum tensor is conserved \cite{LL2,Zwi09}. In multiscale spacetimes, one can adopt the same procedure and end up with a conservation law $\check{D}_\mu\,T^{\mu\nu}_{\rm tot}=0$. This law is not the usual one $\p_\mu\,T^{\mu\nu}=0$, which is not a new effect in varying-$e$ scenarios. In fact, even in the most conservative varying-$\a$ theories such as those pioneered by Bekenstein, the energy-momentum tensor of radiation is not conserved (see, for example, Eq.\ (7) of \cite{SBM}). We will not consider the relativistic particle here, as it would constitute a rather lengthy detour from the main focus of the paper.


\subsection{Electrodynamics}\label{fermions}

We now move to the action of electrodynamics and include fermions,
\bs
\label{qed}\ba
S &=&\int\rmd^Dx\,v\left(\cL_\psi+\cL_m+\cL_F\right)\,,\\
\cL_\psi &=& \rmi\bar\psi\g^\mu D_\mu\psi\,,\\
\cL_m &=&-m\bar\psi\psi\,,
\ea
\es
where $\cL_F$ is given in Eq.\ \Eq{lf}, $\g^\mu$ are the usual Dirac matrices obeying the anticommutation algebra $\{\g^\mu,\g^\nu\}=2\eta^{\mu\nu}$, $\bar\psi:=\psi^\dagger\g^0$ is the Dirac adjoint, and
\be\label{gcd}
D_\mu:= \cD_\mu+\rmi \frac{\tilde e}{\hbar c} A_\mu
\ee
is the gauge covariant derivative. Here we denoted as $\tilde e$ the ``electron charge,'' later to be related to the constant charge $e_0$ and the effective charge $e_v$. Also, the fundamental constants are restored for later comparison with other theories. We still maintain arbitrary $D$ dimensions since we will not consider axial currents, thus ignoring the well-known problem with the pseudoscalar $\gamma_5$ matrix (\cite{Col84}, chapters 4 and 13).

We can formally map the fractional system into one in ordinary spacetime, but with spacetime-dependent effective electric charge. We already discussed the Maxwell term $\cL_F$, and found that the action $S_F$ reduces to the usual one $\cS_\cF$. From the field redefinition\footnote{This formula implies that the fractional spinor representation of the Lorentz algebra follows the same rule of the field representation of \cite{frc6} and of the vector representation above: generators can be obtained from the ordinary ones by multiplying times $1/\sqrt{v}$ to the left and $\sqrt{v}$ to the right.}
\be\label{ppsi}
\Psi:=\sqrt{v}\,\psi\,,
\ee
it is clear that the mass term is $S_m[\psi]=\cS_m[\Psi]=-\int\rmd^Dx\,m\bar\Psi\Psi$. The kinetic Lagrangian $\cL_\psi$ also contains a fermion-gauge interaction,
\ba
S_\psi &=& \rmi\int\rmd^Dx\,v\, \bar\psi\g^\mu \frac{1}{\sqrt{v}}\left(\p_\mu+\rmi \frac{\tilde e}{\hbar c} A_\mu\right)(\sqrt{v}\psi)\nonumber\\
&=& \rmi\int\rmd^Dx\,\bar\Psi\g^\mu \left(\p_\mu+\rmi \frac{\tilde e}{\hbar c} A_\mu\right)\Psi\nonumber\\
&=& \rmi\int\rmd^Dx\,\bar\Psi\g^\mu \left(\p_\mu+\rmi \frac{\tilde e_v}{\hbar c} \cA_\mu\right)\Psi=\cS_\Psi\,,
\ea
where $\tilde e_v=\tilde e/\sqrt{v}$. Here we chose to attach the measure dependence to an effective electron charge rather than to $\hbar$ or the speed of light. However, it would be premature to identify $\tilde e\equiv e_0$ and claim we have recovered Eq.\ \Eq{ev}. Conservations laws will show in a moment, in fact, that
\be
\tilde e_v=e_0\,,
\ee
and that the theory in integer picture is exactly the usual one, $\cS=\cS_\Psi+\cS_m+\cS_\cF$. From now on we reset $c=1=\hbar$.

Thanks to the adoption of the self-adjoint operator $\cD$, the action \Eq{qed} is real-valued. Consider first the standard case without mass and gauge field, i.e., the action $\cS_\Psi$ in the integer picture. Using the properties $(\g^\mu)^\dagger=\g^0\g^\mu\g^0$ and $(\g^0)^2=\mathbbm{1}$, it is easy to show that $\rmi\bar\Psi\g^\mu\p_\mu\Psi-(\rmi\bar\Psi\g^\mu\p_\mu\Psi)^\dagger=\rmi\p_\mu(\bar\Psi\g^\mu\Psi)$. Thus, the kinetic term is self-adjoint up to a total divergence, which can be ignored upon integration. In the fractional case a similar relation holds,
\be\label{ful}
\rmi\bar\psi\g^\mu\cD_\mu\psi-(\rmi\bar\psi\g^\mu\cD_\mu\psi)^\dagger=\rmi\check{\cD}_\mu(\bar\psi\g^\mu\psi)\,,
\ee
and again the last term is a total divergence. This would not have been the case with the derivative of generic weight \Eq{bcD}. Also, the use of ${}_\b\cD$ would determine a Dirac equation which, when ``squared,'' would not yield the Klein--Gordon equation of \cite{frc6}.

The Euler--Lagrange equations,
\ba
&&\cD_\mu\frac{\p\cL}{\p(\cD_\mu\bar\psi)}-\frac{\p\cL}{\p\bar\psi}=-\frac{\p\cL}{\p\bar\psi}=0\,,\nonumber\\
&&\cD_\mu\frac{\p\cL}{\p(\cD_\mu\psi)}-\frac{\p\cL}{\p\psi}=0\,,
\ea
yield the weighted Dirac equation (and its conjugate) with electromagnetic interaction:
\ba
&& \rmi \g^\mu D_\mu\psi-m\psi=0\quad \Rightarrow\quad \rmi \g^\mu\cD_\mu\psi-m\psi=\tilde e A_\mu\g^\mu \psi\,,\nonumber\\\\
&& \rmi D_\mu^\dagger\bar\psi\g^\mu+m\bar\psi=0\quad\Rightarrow\quad \rmi \cD_\mu\bar\psi\g^\mu+m\bar\psi=-\tilde e A_\mu\bar\psi\g^\mu\,.\nonumber\\
\ea
The equation of motion \Eq{eom} for the gauge field yields the Maxwell equation \Eq{max1},
with source
\be\label{edJ}
J^\mu=-\tilde e\bar\psi\g^\mu\psi\,.
\ee
At this point we can determine $\tilde e$. The conservation law \Eq{dj} should be compared with Eq.\ \Eq{ful}, $\check{\cD}_\mu (J^\mu/\tilde e)=0$. Since $\check{\cD}=v^{-1/2} \cD [v^{1/2}\,\cdot\,]$, Eqs.\ \Eq{dj} and \Eq{ful} are compatible if, and only if,
\be\label{tile}
\boxd{\tilde e =\sqrt{v}\,e_0\,.}
\ee
This is consistent also with Eq.\ \Eq{FcF2}, since
\ba
\fl J^\mu &=& -\sqrt{v}\,e_0\bar\psi\g^\mu\psi= -\frac{\sqrt{v}\,e_0}{v}(\sqrt{v}\bar\psi)\g^\mu(\sqrt{v}\psi)\nonumber\\
 &=&-\frac{e_0}{\sqrt{v}}\,\bar\Psi\g^\mu\Psi =\frac{\cJ^\mu}{\sqrt{v}}\,.
\ea
 
The improved energy-momentum tensor is the generalization of Eq.\ \Eq{emt},
\ba
\fl T_{\mu\s} &:=&\eta_{\mu\s}\cL-\frac{\p \cL}{\p (\cD^\mu A_\nu)}\cD_\s A_\nu-\frac{\p\cL}{\p\cD^\mu\bar\psi}\cD_\s\bar\psi\nonumber\\
&&-\frac{\p\cL}{\p\cD^\mu\psi}\cD_\s\psi+\check{\cD}_\nu(F^\nu_{\ \mu} A_\s)\nonumber\\
\fl &=& -\frac{1}{4}F^{\nu\tau} F_{\nu\tau}\eta_{\mu\s} +F_\mu^{\ \nu}F_{\s\nu}-\rmi\bar\psi\g_\mu D_\s\psi\,,\label{emt3}
\ea
where we used the fact that $\cL_\psi+\cL_m=0$ on shell and the added correcting term (total weight-1 derivative in the first line) completes both the second term in \Eq{emt3} into a gauge-invariant expression and the derivative in the last term into a gauge covariant derivative \Eq{gcd}.


\section{Comparison with models with varying couplings}\label{comp}

In the introduction, we mentioned the existence of various scenarios where the fine-structure constant $\alpha$ depends on the spacetime point. These models may or may not break local Lorentz and diffeomorphism invariance, but we shall assume that they do not. Even then, we have a choice between varying-$e$ and varying-$c$ models, with different phenomenological implications. Here we attempt to set up a bridge between our construction and these models. In them, $\alpha$ is dynamically determined by a scalar field, in contrast with the results presented here. Even putting aside this point, the theories are structurally quite distinct in the form a varying coupling appears. Still, we find several points of contact between these constructions.

\subsection{Varying electron charge}

A possibility for varying $\alpha$ is to keep $\hbar$ and $c$ constant, but allow for a nonconstant electric charge:
\be
e_0\to e(x)\,.
\ee
Bekenstein proposed a phenomenological varying-$e$ model based on some reasonable assumptions: standard Maxwell equations, dynamical origin of variations of $\a$, validity of action principle, gauge invariance, time-reversal invariance, causality (absence of ghosts), Planck length being the shortest physical scale, and Einstein's equations \cite{Bek82}. The resulting Maxwell action is of the form
\ba
S_F &=& -\frac14\int\rmd^4x\,\sqrt{-g}\, F_{\mu\nu} F^{\mu\nu}\,,\nonumber\\
F_{\mu\nu} &:=&\frac{1}{e(x)}\{\p_\mu[e(x)A_\nu]-\p_\nu[e(x)A_\mu]\}\,,\label{bekac}
\ea
where $g$ is the determinant of the metric and $e(x)$ is governed by the dynamics \cite{Bek82,BaO,MSK,Bek02} via the action 
\be
S_e:=-\frac{\hbar c}{2 l^2}\int\rmd^4x\,\sqrt{-g}\,e^{-2}\p_\mu e \p^\mu e\,,
\ee
where $l$ is some length assumed to be larger than the Planck scale $\lp=\sqrt{\hbar G/c^3}$ and smaller than the scales where standard electromagnetism is verified to high accuracy, $\lp\leq l<10^{-17}$m. The theory can be recast as a dilatonlike model where only the electromagnetic Lagrangian is nonminimally coupled to the scalar field $\Phi:=\ln (e/e_0)$ \cite{MSK}. By defining $a_\mu:=(e/e_0) A_\mu$ and $f_{\mu\nu}:=(e/e_0) F_{\mu\nu}=\p_\mu a_\nu-\p_\nu a_\mu$, the total action reads
\be\label{ace}
S=\int\rmd^4x\,\sqrt{-g}\,\left(\cL_g+\cL_{\rm mat}+\rme^{-2\Phi}\cL_f -\frac{\om_e}{2}\p_\mu \Phi \p^\mu \Phi\right)\,,
\ee
where $\cL_g\propto R$ is the gravity Lagrangian ($R$ is the Ricci scalar), $\cL_{\rm mat}$ is the contribution of all the other matter components, $\cL_f=-(1/4)f_{\mu\nu} f^{\mu\nu}$, and $\om_e=\hbar c/l^2$. Notice that $\Phi$ does not appear inside $\cL_{\rm mat}$, once the coupling between fermions and the Abelian gauge field is rewritten in terms of the new variable.

Astrophysical and purely electromagnetic experiments are not able to rule out a time-varying $\a$ \cite{Bek82}, nor do tests of the weak equivalence principle \cite{Bek02}.\footnote{From a cursory inspection of Eq.\ \Eq{ace}, one might speculate that varying-$e$ models violate the principle since the scalar $\Phi$ couples differently to different matter species. However, it turns out that the anomalous electric force acting on a charged particle with mass $m(\Phi)$ is canceled by corrections to the usual Coulomb force. Thus, different particles with same charge-to-mass ratio experience the same acceleration in external fields, and the weak equivalence principle is respected \cite{Bek02}.} Cosmology \cite{Bek02,SBM,BSM,OP1,BMS,MoB} is compatible with the allowed variation of the fine-structure constant during the evolution of the Universe, including at redshifts $z\lesssim 4$ as per Eq.\
\Eq{deaa} \cite{Mur03}. A varying $\a$ has consequences for the whole Standard Model of particles. If the electric charge varies in spacetime, one expects that, at high energies, all the gauge couplings of the electroweak sector are scalar fields \cite{KiM,ShB1}, leading to variable gauge-boson masses but constant lepton masses after spontaneous symmetry breaking of $SU(2)_L\times U(1)_Y\to U(1)_{\rm EM}$. In the QCD sector, an extension of Bekenstein's theory to a varying strong-coupling parameter is excluded experimentally \cite{CLV}, while in grand unification scenarios one obtains that the nucleon mass, the magnetic moment of the nucleon, and the weak coupling constant all vary \cite{CaF,LSS,DNRV}. 

Also, assuming that the effective electric charge is embodied by a Lorentz scalar field, as in Bekenstein's model, changes in $\a$ would lead to tremendous changes in the vacuum energy, which can be made compatible with observations only via large fine tunings \cite{BDD}.

How does this model compare with the one proposed here? Most notably, our varying $\alpha$ is not driven by a scalar field, but is prefixed in spacetime due to its anomalous geometric structure. Therefore, the phenomenological constraints mentioned above are simply not applicable to our work. But even setting aside this very important practical detail (which does not hold in certain versions of multiscale spacetimes where the measure is taken as a dynamical field \cite{fra1,fra2}), there is a crucial difference. The new variables $a_\mu$ and $f_{\mu\nu}$ here look like the ``integer picture'' variables ${\cal A}_\mu$ and ${\cal F}_{\mu\nu}$, but when we look at the action in terms of the new variables we note a fundamental change. In Bekenstein-style theories, a dilaton coupling $\rme^{-2\Phi}$ appears in front of the ``$F^2$'' Maxwell term, unlike in the theory presented here. Furthermore, $\Phi$ disappears from the couplings to other matter components, in contrast with interacting multiscale theories in general \cite{frc6}.

For these reasons, we can conclude that the theory presented in this paper can {\it not} be reduced to the previously proposed varying $e$ reviewed in this section, both dynamically and structurally.


\subsection{Varying speed of light and Planck constant}

In other scenarios, the speed of light is made to be spacetime dependent \cite{Mag03}:
\be\label{vsl}
c_0\to c(x)\,.
\ee
A varying speed of light (VSL) was early recognized as having an impact in the history of the universe \cite{Mof92,AlM,BaM1,BaM2,BaM3,Bar99}. The cosmological applications of Eq.\
 \Eq{vsl} may have some problems \cite{LSV,ElU}, but these can be overcome in a different incarnation of \Eq{vsl} \cite{Mag03,Mag00,MBS}. More recently, VSL models reemerged as strong contenders to explain the observed cosmic structure \cite{Mag08} both in a bimetric reformulation \cite{Mag08b,Mag10} and in the context of deformed dispersion relations~\cite{Mag08c}.

Not only does the interest in a varying speed of light lie in its intimate relation to varying-$e$ models \cite{BaM1,MBS}, but also in the fact that VSL theories simply correspond to frameworks where units are adapted with the scales in the dynamics \cite{Mag00} (and, in particular, chosen such that $c$ varies). Time and space units are redefined so that the differentials scale as $\rmd t\to [f(x)]^a \rmd t$, $\rmd x^i\to [f(x)]^b \rmd x^i$, where $f$ is a function, $a$ and $b$ are constants, and local Lorentz invariance of the line element requires $c(x)\propto [f(x)]^{b-a}$. We recognize here a particular form of anisotropic multiscaling (one that distinguishes between space and time variables). 
In particular, when $b=0$ one formally reabsorbs $c$ in the coordinate
\be\label{x0vsl}
x^0=\int\rmd t\, c(t)\,,
\ee
which scales as a length. With this coordinate, all equations can be made general-covariant and gauge invariant (in a word, formally identical to the usual ones) provided some conditions are met. For instance, the field strength of the Abelian electromagnetic field $A$ is of the form
\be
F_{\mu\nu}=\frac{\hbar c}{e}\left[\p_\mu\left(\frac{e}{\hbar c}A_\nu\right)-\p_\nu\left(\frac{e}{\hbar c}A_\mu\right)\right]\,,
\ee
and explicit $c$ dependence [hence, coordinate dependence, unlike Eq.\ \Eq{bekac}] disappears if $e\propto \hbar c$ (a major difference with respect to the theory proposed in this paper). However, even with this constraint, we may still have a varying $\alpha$. Specifically, if  
\be\label{cea}
e\propto \hbar(c) c\propto c^q\,,
\ee
then
\be\label{cea1}
\a\propto \hbar c \propto c^q\, ,
\ee
and, as long as $q\neq 0$, we have nontrivial effects. Notice that this amounts to a statement on Planck's constant:
\be
\hbar(c) \propto c^{q-1}\,.
\ee
Moreover, the gravitational dynamics is nontrivial.
VSL models are locally Lorentz invariant (namely, under transformations which look like the usual ones but with $c_0$ replaced by a varying $c$).
Defining the scalar field $\chi:=\ln (c/c_0)$, the total action for a minimal version of VSL can be written as \cite{Mag00}
\be\label{acc}
S=\int\rmd^4x\,\sqrt{-g}\,\left[\cL_g+\rme^{b\chi}\left(\cL_{\rm mat}+\cL_F\right)-\frac{\om_c}{2}\p_\mu \chi \p^\mu \chi\right],
\ee
where $b$ and $\om_c$ are constants. Only when $q=0$ is the theory equivalent to a Brans--Dicke model \cite{Mag00}.

Models where $e$ or $c$ varies can be recast in new units such that, respectively, the electric charge and the speed of light become constant, but in both cases the dynamics become substantially more complicated. This criterion of simplicity is not the only one which attaches one label or the other (varying-$e$ versus varying-$c$) to these models: experiments are able to distinguish between them. Concerning spatial variations, in VSL $\a$ increases near compact objects (such as black holes) and electromagnetism may become nonperturbative, while in varying-$e$ models the opposite happens and $\a$ decreases \cite{MBS}; this may lead to distinctive predictions for the cosmic microwave background temperature and polarization spectra \cite{SKK}. Furthermore, tests of the weak equivalence principle can distinguish between the two classes of models, since whereas VSL theories abide by this principle, Bekenstein's model does not \cite{MBS}. Also, the sign of the time variation of $\a$ is in agreement with Keck observations \cite{MWF,Mur03} in both theories. VSL scenarios are generally able to predict a decrease in $\a$ in the past for all types of dark matter \cite{MBS}, but so does the varying-$e$ theory, should the dark matter electromagnetic energy be of the magnetic type  \cite{Bek02}.

Once again, none of these constraints applies to the model proposed in this paper. However, there are obvious structural similarities between them, should we be prepared to contemplate anisotropic scaling dimensions (differentiation between space and time) \cite{ACOS,fra7}.


\subsection{Comparison with multiscale spacetimes}
Here we list in a more systematic way the differences between the varying $\alpha$ models previously proposed and ours. 
\begin{enumerate}
\item Multifractional theory and the varying-$e$ proposal are invariant under deformed gauge transformations such that the dependence on the effective electron charge is inverted. In the multifractional scenario this transformation is Eq.\ \Eq{gauge}, which can be also written equivalently as $e_v^{-1}A_\mu\to e_v^{-1}A_\mu+\p_\mu\vp$, or $e_0^{-1}\cA_\mu\to e_0^{-1}\cA_\mu+\p_\mu\vp$, where $\vp=\phi/e_v$ is an ordinary scalar. On the other hand, Bekenstein's theory is invariant under $e A_\mu\to e A_\mu+\p_\mu\vp$, where $e=e(x)$.
The correspondence $e_v\leftrightarrow e^{-1}$ is also apparent in the comparison of Eqs.\ \Eq{myF} and \Eq{bekac}. Still, we can play with Eq.\ \Eq{myF} and make it formally equivalent to \Eq{bekac} under the replacement $\p\leftrightarrow \check{\cD}$. In fact, noting that $v e_v = e_0^2/e_v$, the multifractional Maxwell field strength is $F_{\mu\nu}= e_v^{-1}[\check{\cD}_\mu(e_v A_\nu)-\check{\cD}_\nu(e_v A_\mu)]$.
\item In our case, the variation of the fine-structure constant stems from a variation of $e\to e_*(t)$ of purely geometric nature. In the absence of dynamics for geometry (both for metric and measure structures), this time dependence is nondynamical, contrary to both varying-$e$ and varying-$c$ models, and $e_*(t)$ is a given profile motivated by multifractal geometry. A consequence of this (further discussed in the final section) is that the variation law of the fine-structure constant does not change during the history of the Universe, contrary to the cosmology of Bekenstein's model \cite{SBM,BSM,OP1,BMS,MoB}.
\item In the VSL context, it is intriguing to notice how the change of units at the base of that proposal maps into the notion of ``adapting rods'' in multiscale models \cite{fra7}. Equation \Eq{x0vsl} corresponds to a coordinate redefinition such that the theory action looks trivial from the point of view of general covariance and Lorentz invariance, but in fact it hides a unit redefinition of the coordinates. This, in turn, corresponds to a highly nontrivial choice of momentum space \cite{frc7}. In multifractional theories exactly the same thing happens: the units of the coordinates are defined in a way making geometry (in particular, momentum space) nontrivial and anomalous, although it is possible to formally (and only to some extent) map the theory into a usual one with some modifications. The resemblance with VSL models is especially striking for the multifractional theory with $q$-Laplacian \cite{frc2,fra7,frc7}, where the most general action in position space can be made identical to the usual one when using anomalous coordinates $q(x)$ (scaling as $[q^\mu]\sim-\a_\mu$ in various regimes) which actually depend on coordinates $x$ with normal scaling ($[x^\mu]=-1$). Anomalous coordinates feature also in Ho\v{r}ava--Lifshitz gravity \cite{Hor2,Hor3}, where, however, the scaling is implicit and coordinates are not composite objects. They can be made composite by mapping the model to a multifractional one with anisotropic measure \cite{fra7}; this connection (valid up to the choice of symmetries, which crucially determine the form of the Laplacian in either theory) is based upon the presence of a hierarchy of scales, either hidden or explicit. For the very same reason, the geometries of multifractional, Ho\v{r}ava--Lifshitz, and VSL scenarios share several similarities.
\item A comment is in order on the resemblance of the multifractional measure weight with a dilaton or a Brans--Dicke scenario. The dilaton of string theory couples differently in different matter sectors, but nonperturbative effects may render its coupling universal \cite{DaP} (see also \cite{OP1,DvZ,OP2}). This is somewhat similar to the fractal model of \cite{fra1,fra2}, where $v$ was regarded as a scalar field appearing as a global rescaling of the standard Lagrangian. As we saw above, also varying-$e$ and VSL theories feature a nonminimal coupling between matter and Maxwell sectors and a scalar field, Eqs.\ \Eq{ace} and \Eq{acc}. On the other hand, in the framework studied in the present paper the measure weight is not a Lorentz scalar but a fixed coordinate profile without kinetic term (whose shape is dictated by fractal geometry rather precisely \cite{frc1,frc2}), changing the differential structure of the geometry. The presence of a nontrivial measure consistently affects the definition of functional variations, Poisson brackets and Dirac distribution, in turn leading to a deformation of the Poincar\'e symmetries \cite{frc6}. On the other hand, Lorentz invariance is respected in both varying-$e$ and varying-$c$, although Lorentz matrices are themselves ``deformed'' in the second case.
\item A consequence of the last point is that the argument of \cite{BDD} severely against appreciable variations of $\a$ may be avoided in multifractional field theories. In fact, $\a(x)$ is not a scalar field but a geometry-dependent object, and one would be entitled to subtract off the vacuum-energy shift in the cosmological constant for all values of $\a$ (if $\a$ were a scalar, the subtraction would proceed only for its value at low energy or late times).
\item Even reverting to the interpretation of \cite{fra1,fra2} of $v$ as a scalar and reinstating a kinetic term (and risking to incur in the objection of \cite{BDD}), the multiscale scenario would be significantly different from the others. In the fundamental formulation with field densities, the weight $v$ modifies all derivatives in the action, in all sectors. Even when translating the theory from field densities to fields, the gravitational sector does not become trivial, since $\cL_g$ is highly nonlinear in the metric \cite{frc11}. Furthermore, Poincar\'e symmetries would be deformed in the characteristic way of anomalous geometries, as explained above \cite{frc6}. Therefore, multiscale spacetimes cannot be made physically equivalent to any of the above dilatonlike proposals.
\item As a bird's eye view, one can state that multifractional spacetimes possess a mixture of properties which make them akin to varying-$e$ and varying-$c$ models in different ways. On one hand, by construction an effective spacetime-dependent electric charge naturally arises in multiscale spacetimes. On the other hand, the latter always predict a decrease of $\a$ in the past regardless the matter content, just like VSL \cite{MBS} and, in a subtler way, varying-$e$ models \cite{Bek02}. Also, in multiscale theories the spacetime profile of all effectively varying couplings only depend on the measure weight $v(x)$, just as the couplings in VSL models depend on the profile \Eq{vsl} alone, Eq.\ \Eq{acc}. In this sense, multiscale spacetimes respect the weak equivalence principle as VSL and varying-$e$ models, although we have not really discussed gravity in the present context.
\end{enumerate}


\section{Discussion}

In this paper, we have worked out the field theory for electrodynamics in a particular class of multiscale spacetimes. Fermions and the $U(1)$ gauge field have been introduced extending previous knowledge of field theories living in such spacetimes, which was limited to real scalars \cite{frc6}. This opens up the possibility to study non-Abelian gauge fields and, in particular, the electroweak Standard Model. A discussion of the latter along the same lines of \cite{KiM} will be left for the future.

In general, spacetime-dependent couplings naturally arise in this framework. Electrodynamics in multiscale geometry displays a varying electric charge whose profile is determined by the measure, thus falling into a class of models whose phenomenology has been extensively discussed in the literature. The type of coordinate dependence of the charge, however, as well as its motivation and the overall theoretical structure, widely differ with respect to other varying-$e$ models, as detailed above. It also makes this model distinct from the covariant varying-$c$ scenarios previously proposed. In spite of various similarities, the present one is a genuinely new proposal for a varying-constants theory. 

A remark valid for both the present multiscale theory and previous proposals is in order. One should note the difference between the conserved electric charge, for example in Bekenstein's theory, and the charge that actually couples to electromagnetism. The number $n$ of electrons is conserved even when dilaton field $\Phi$ varies, so $ne_0$ obviously provides a conserved charge, but this is not what couples to the electromagnetic sector. This has some implications when applying experimental constraints for the conserved charge which is, by definition, conserved even in varying-$e$ (and also varying-$c$) theories. Therefore, one should carefully examine the type of experiment and theoretical observable in order to relate one to the other in a nonmisleading way. In the case of the theory in multiscale spacetime, there is a further complication. We have found two types of charges, the one appearing in the conservation law of the Noether current [$Q$ and $e_*$, Eqs.\ \Eq{qch} and \Eq{e*}] and the one coupling electrons to the electromagnetic field [$\tilde e(x)$, Eq.\ \Eq{tile}]. It is the former, in fact, that we should compare with experiments, since $\tilde e$ is just a quantity appearing in the action which, when properly manipulated, leads consistently to $e_v$. 

As we stated before, none of the phenomenological constraints valid for previously proposed varying-$\alpha$ theories can be applied directly to our work. We conclude with an estimate of the variation of the fine-structure constant in multiscale spacetimes, between a time $t$ in the past and today ($t_0$). From Eq.\ \Eq{e*}, it follows that
\ba
\frac{\Delta\a}{\a}&=&\frac{v_*(t_0)}{v_*(t)}-1=-\frac{\left|\frac{t_*}{t}\right|^{1-\a_0}-\left|\frac{t_*}{t_0}\right|^{1-\a_0}}{1+\left|\frac{t_*}{t}\right|^{1-\a_0}}\nonumber\\
&\approx& -\frac{1}{1+\left|\frac{t}{t_*}\right|^{1-\a_0}},
\ea
where in the last step we assumed that $t_0\gg t_*$.\footnote{Multiscale theories are not translation invariant in the usual sense because the measure fixes a frame. The background dependence establishes that there is an origin which, in the time direction, we assumed in the text to coincide with the big bang. On the other hand, in another version of the theory where the measure is $v(x-x')$ instead of $v(x)$, the measure singularity is at some nonzero ${x'}^\mu$. The geometry, however, does not change: spectral and Hausdorff dimension will remain the same, since one has not changed the scaling law of $v$. In this respect, the translation of the origin can be regarded as a different {presentation} of the theory. However, when comparing with experiments as in the text, one will notice this $x^\mu-{x'}^\mu$ shift in the measure and will have to state, for instance, what $t'$ is in relation to the history of the Universe.} Notice that $\Delta\a/\a<0$, in intriguing agreement with the signature of the effect \Eq{deaa} in quasar observations at Keck \cite{Mur03} [but in disagreement with VLT results \cite{Web10,Kin12}, Eq.\ \Eq{deaa2}]. If $t_*$ was Planck time, the corrective effect would be completely negligible. Taking instead $\a_0=1/2$ (a reasonable value at small scales or early times \cite{frc1,frc2}) and the observational bound \Eq{deaa} applied to $t\sim 1.79$\,Gyr ($z\sim 3.5$, using the best-fit values of the parameters of \cite{Pl13a}) from the big bang, we obtain an estimate for the intrinsic time scale of the measure:
\be\label{t*}
t_*\sim 0.058~{\rm yr}\,,
\ee
stating that multifractional effects on the geometry of the universe have become negligible after about 21 days since the big bang (which we assumed to be at $t=0$). We can plug back this value to estimate $\Delta\a/\a$ at other times. For the small-redshift quasars ($t\sim 10$\,Gyr), we get $\Delta\a/\a\approx-2.4\times 10^{-6}$, quite compatible with the bound $|\Delta\a/\a|<10^{-5}$ of \cite{DWBF,Car00,Mu01a}.\footnote{For the Oklo natural reactor ($t\sim 12$\,Gyr), we obtain $\Delta\a/\a\approx-2.2\times 10^{-6}$, about two orders of magnitude larger than the present constraints \cite{Uza10}. However, this bound is strongly model dependent and possibly subject to criticism \cite{CaF,LSS}, and it should be taken \emph{cum grano salis}.} Extrapolating back to big-bang nucleosynthesis (BBN, $t\sim 2~{\rm s}\div 20~{\rm min}$) leads to an effect of order of $\Delta\a/\a\sim -0.98$, which is excluded experimentally (standard BBN can tolerate as much as $O(10^{-2})$ variations \cite{Uza10,Ave01}). This may be interpreted either by regarding the estimate \Eq{t*} as too large (but in this case varying $\a$ in quasar observations would not be explained by the multifractional model; matching allowed $\a$-variations during BBN yields $t_*<0.3$\,s), or as a failure of real-order multifractional measures and the appearance of a finer hierarchy of scales in the very early Universe \cite{frc2,ACOS}, or, again, as the effect of naive simplifications such as ignoring gravity. In particular, adopting a polynomial rather than a binomial measure \cite{frc4} might account for a more resilient history of the Universe and a better fit to datasets. Also, inclusion of the full spatial dependence of the measure (i.e., probing smaller spatial scales, which are amplified to cosmological size during the history of the early universe) might interestingly confront the theory with the preferred-frame or ``dipole'' effect allegedly found in quasars data \cite{Web10,Kin12} (see \cite{OPU} for an explanation in dilatonlike theories). Further study will hopefully refine the multiscale model and its predictions.


\section*{Acknowledgments}
We thank Luis J.\ Garay and Giuseppe Nardelli for useful discussions. The work of G.C.\ is under a Ram\'on y Cajal contract. J.M.\ was funded by STFC through a consolidated grant and by an International Exchange Grant from the Royal Society.




\end{document}